\documentclass[a4paper,11pt]{article}
\usepackage{jheppub}
\usepackage[utf8]{inputenc}
\usepackage{mathtools}
\usepackage{listings}
\usepackage{tikz}
\usepackage{amsfonts}
\usepackage{amsmath}
\usetikzlibrary{arrows.meta}

\usepackage{pifont}% http://ctan.org/pkg/pifont
\usepackage{dsfont} 

\usepackage{comment}

\graphicspath{{figures/}}	 % set directory for figures
\usepackage{tikz}
\usetikzlibrary{arrows,calc,snakes,shapes,shapes.arrows,
decorations.markings,decorations.pathmorphing}
\newcounter{sarrow}

     \tikzset{>=triangle 45}
     \tikzstyle{gr}=[draw,circle,green!50!black,fill=green!50!black,scale=.6]
     \tikzstyle{Bl}=[draw,circle,blue,scale=.7]
     \tikzstyle{R}=[draw,circle,fill=red,scale=.7]
     \tikzstyle{bl}=[draw,circle,fill=black,scale=.2]
     \tikzstyle{bbc}=[draw,circle,fill=black,scale=.75]
 
\usepackage{array} % ??
\usepackage{multirow} % multiple row elements in a table
\usepackage{colortbl} % colored rows and columns in tables
\usepackage{arydshln} % dashed lines in arrays and tables
\usepackage{stfloats}  % ??
\usepackage{cellspace}
\newcommand{\xdasharrow}[2][->]{
\tikz[baseline=-\the\dimexpr\fontdimen22\textfont2\relax]{
\node[anchor=south,font=\scriptsize, inner ysep=1.5pt,outer xsep=2.2pt](x){#2};
\draw[shorten <=3.4pt,shorten >=3.4pt,dashed,#1](x.south west)--(x.south east);
}}
\usepackage{hhline} % to keep border in colored cells

\usepackage{xcolor}   % \color{...}, colored text
\def\Sf{\mathfrak{S}}

\newcommand{\beq}{\begin{equation}}
\newcommand{\eeq}{\end{equation}}
\newcommand{\bpm}{\begin{pmatrix}}
\newcommand{\epm}{\end{pmatrix}}
\newcommand{\bpmat}{\begin{pmatrix}}
\newcommand{\epmat}{\end{pmatrix}}
\newcommand{\bsmat}{\begin{smallmatrix}}
\newcommand{\esmat}{\end{smallmatrix}}

\def\^{\wedge}

\def\sof{\mathop{\mathfrak{so}}}

\def\C{\mathbb{C}}

\def\Z{\mathbb{Z}}

\def\cM{{\mathcal M}}
\def\cN{{\mathcal N}}

\def\cT{{\mathcal T}}

\def\-{{\text{-}}}

\title{Snowmass White Paper on SCFTs  }

\author[1]{Philip C. Argyres,}
\author[2]{Jonathan J. Heckman,}
\author[3]{\\Kenneth Intriligator,}
\author[4,5,6]{and Mario Martone}
\affiliation[1]{Physics Department, University of Cincinnati, Cincinnati, OH 45221, USA}
\affiliation[2]{Department of Physics and Astronomy, University of Pennsylvania, Philadelphia, PA 19104, USA}
\affiliation[3]{Department of Physics, University of California, San Diego, La Jolla, CA 92093, USA}
\affiliation[4]{Simons Center for Geometry and Physics, Stony Brook University,

Stony Brook, NY 11794-3840, USA}
\affiliation[5]{C.~N.~Yang Institute for Theoretical Physics,  Stony Brook University,

Stony Brook, NY 11794-3840, USA}
\affiliation[6]{Department of Mathematics, King's College London, The Strand, London WC2R 2LS, U.K.}
\emailAdd{\\philip.argyres@uc.edu}
\emailAdd{\\jheckman@sas.upenn.edu}
\emailAdd{\\keni@ucsd.edu}
\emailAdd{\\mario.martone@kcl.ac.uk}
\notoc 
\abstract{Superconformal field theories (SCFTs) occupy a central role in the study of many aspects of 
quantum field theory. In this white paper for the Snowmass process we give a brief overview of aspects of SCFTs in $3\leq D \leq 6$ space-time dimensions, including classification 
efforts and some of the vast current research trends on the physical and mathematical 
structures generated by this rich class of physical theories.}

\begin{document}

\maketitle

\section{Background and Motivation}

A key problem in theoretical physics is to gain an analytical understanding of the dynamics of strongly-coupled quantum field theories.  
This understanding is needed to inform and constrain both the search for effective theories which go beyond the Standard Model of particle physics, and the search for effective descriptions of new collective behaviors in condensed matter systems.  
Among quantum field theories, conformal field theories (CFTs) play a central role.  
As possible UV and IR endpoints of renormalization group (RG) flows they provide launching points, via deformations, for exploring the broader space of quantum field theories.  
In addition, via the AdS/CFT correspondence, CFTs encode aspects of space-time in quantum gravity, and so their study may also illuminate physics outside of the framework of local quantum field theory.

Without any further assumptions beyond conformal invariance, locality, and unitarity, a systematic understanding of the landscape of CFTs remains largely out of reach.   
The $\sof(D, 2)$ conformal symmetry algebra together with unitarity implies useful but weak constraints on the allowed spins and dimensions of local operators in a CFT; see \cite{Mack:1975je, Minwalla:1997ka, Dolan:2005wy, Kinney:2005ej, Maldacena:2011jn,Rychkov:2016iqz} and references therein.  We will here limit our discussion to $D>2$ spacetime dimensions. Following the landmark BPZ paper~\cite{Belavin:1984vu}, many thousands of papers explored the rich realm of RG flows and CFTs in $D=2$ spacetime dimensions and the special power of the enhancement of $\sof(D, 2)$ to the left and right Virasoro algebras; the subject of $D=2$ is too rich, and too special to include here.   Likewise, the $AdS_{D+1}/CFT_D$ correspondence is too rich and vast to attempt to be properly represented in this brief overview, see e.g. \cite{Aharony:1999ti}.

As we increase $D$, interactions are IR-weaker and UV-stronger, so interacting CFTs become less likely, and more challenging to find and analyze.  In space-time dimensions $D=3$ and $D=4$ there are a few classes of CFTs which are perturbatively accessible (close to free field theory) \cite{Belavin:1974gu, Caswell:1974gg, Banks:1981nn}.  There are also examples of perturbatively accessible asymptotically safe UV CFT fixed points~\cite{Litim:2014uca}.  $D=4$ non-Abelian gauge theories have a conformal window of matter content that flows to a CFT in the IR, where the lower limit of the conformal window is a strong coupling question that is debated on a case-by-case basis in the lattice gauge theory community, see e.g.~\cite{Fodor:2018uih} and references therein. 
%whose IR divergences are not under control.  
The bootstrap program \cite{Bootstrap}, based on demanding the associativity of the product of local operators, gives numerical constraints on the existence of unitary CFTs with low values of conformal central charges (vaguely, a measure of the number of degrees of freedom or complexity of the CFT) but becomes increasingly difficult to implement as the central charges increase.
The large $N$ or large central charge limit of generic CFTs is generally challenging.

The existence and tractability of CFTs improves dramatically with the introduction of supersymmetry.  Supersymmetry gives exact analytic control over a large set of important observables in supersymmetric QFTs, and the constraints and methods are profoundly extended for superconformal QFTs (SCFTs).  The existence of SCFTs in $D=4$ \cite{Nahm:1977tg,Sohnius:1981sn,Brink:1982wv,Grisaru:1982zh,Howe:1983wj,Parkes:1984dh,West:1984dg,Parkes:1985hj} and other $D$ have thus attracted considerable attention from the high energy theory community over the past few decades,
%25 years, 
and has grown into a mature sub-field; see e.g.~\cite{Intriligator:1995au,Chaichian:2000wr,Tachikawa:2018sae,Akhond:2021xio} for some reviews and references. For superconformal theories, $\sof(D, 2)$ is part of a superalgebra $\Sf$ whose fermionic supercharges are in the spinor representation. 
This is highly restrictive and only possible for $D\leq 6$ \cite{Nahm:1977tg}. 
$\Sf$ also contains a bosonic $R$-symmetry subalgebra which is an internal global symmetry of the SCFT.  
The constraints on operator content coming from combining superconformal invariance with unitarity are substantially strengthened; see e.g. \cite{Minwalla:1997ka, Dolan:2002zh, Cordova:2016emh} and references therein.   SCFTs in $D>2$ moreover typically have moduli spaces of supersymmetric vacua, where the theory becomes more weakly coupled, which greatly aid in their analysis.

In $D=3$ and $D=4$ (but not $D>4$), there are many classes of SCFTs that come in continuous families, called ``conformal manifolds", labeled by dimensionless couplings for exactly marginal operators, and there can be strong-weak coupling dualities on such spaces. In many such classes, the conformal manifold has weakly-coupled regions, where perturbation theory can be applied \cite{Drummond:2006rz, Arkani-Hamed:2009pfk, Arkani-Hamed:2009kmp, Arkani-Hamed:2010zjl, Beisert:2010jr, Bourjaily:2011hi, Arkani-Hamed:2012zlh, Bourjaily:2013mma, Arkani-Hamed:2013jha, Bourjaily:2015jna, Bourjaily:2015bpz, Bourjaily:2016evz, Dixon:2016nkn, Bourjaily:2017wjl, Drummond:2018caf, Caron-Huot:2019vjl, Caron-Huot:2019bsq}.  In dimensions 3, 4 and 6 and with enough supersymmetry, infinite subalgebras of the operator product algebra form 1D topological algebras or vertex operator algebras (VOAs) whose structure is tightly constrained and can be computed in many examples; see \cite{Lemos:2020pqv} and references therein.
Furthermore, knowledge of these subalgebras greatly enhances the reach of bootstrap techniques; see \cite{Bootstrap}.
The large $N$ or large central charge limits of many classes of SCFTs can be understood using semi-classical supergravity or superstring techniques via the AdS/CFT correspondence.

For $D>4$, all Lagrangian theories are IR free, so interacting SCFTs appear unlikely from a traditional QFT perspective. Nevertheless,  infinite classes of strongly-coupled SCFTs in $D=5$ and $D=6$ dimensions have been constructed and explored by string / M- / F- theory compactifications or brane localizations which decouple gravity and preserve supersymmetry.  The existence of $D>4$ interacting SCFTs is a guide to developing deeper perspectives, and new methods, for understanding QFTs in all dimensions.  The $D>4$ SCFTs are interesting in their own right, and have also served as a gateway to understand and organize a wide variety of lower-dimensional strongly coupled phenomena.

Starting with 6D SCFTs as ``master theories'' one obtains field theories in $6-k$ dimensions by compactification on a $k$-dimensional manifold $X_k$, relating in this way the properties of the lower dimensional theories with the geometric properties of $X_k$. 
Perhaps the best known example is the 2-torus, $T^2$, compactification of 6D $\cN=(2,0)$ SCFTs, which yields 4D theories with sixteen supercharges, namely $\cN = 4$ super Yang-Mills theories.
These 4D $\cN=4$ gauge theories are conjectured to be S-dual \cite{Goddard:1976qe, Osborn:1979tq} --- their infinitely-strong coupling limits are equivalent to their weakly-coupled limits --- a conjecture which seems too difficult to prove using 4D field theory techniques as it involves infinite-coupling limits.  
But from the 6D perspective, the complex structure of the $T^2$ is the holomorphic gauge coupling $\tau$ of the 4D theory, and S-duality is evident geometrically as the invariance  of the complex structure of the torus under $\tau \to -1/\tau$ \cite{Vafa:1997mh}.
Generalizations include the study of such 6D theories on other Riemann surfaces besides $T^2$, including the possibility of punctures \cite{Seiberg:1994aj, Seiberg:1994rs, Argyres:1995jj, Argyres:1995xn, Argyres:1996eh, Gaiotto:2009we, Gaiotto:2009hg, Argyres:2007cn}, where  global symmetries of the moduli space of the compactification again translate to highly non-trivial duality transformations in the 4D effective SCFT. 

Similar insights have followed for compactifications on other spaces, including singular ones, leading to an illuminating correspondence between the structure of higher-dimensional theories and their lower-dimensional counterparts. 
This perspective has also been used to understand extended operators and higher-form symmetries of lower dimensional theories.
Moreover, the correspondence between higher- and lower-dimensional SCFTs can be used ``in reverse" to inform the mathematical study of the geometry of singular spaces.  See \cite{PhysMat} for more about the wonderful synergy between QFT and mathematics, which has led to important new insights in both directions.

\paragraph{Organization} 

In the rest of this white paper we provide a summary of the current state of knowledge and directions of active research.
Section \ref{sec:classification} provides an overview of the status of the systematic classification of SCFTs for each dimension starting from $D=6$ to $D=3$. 
Section \ref{sec:newMath} summarizes the numerous mathematical structures which the study of SCFTs has helped in understanding; we focus specifically on the implications for geometry, representation theory, and vertex operator algebras. 
Section \ref{sec:extOp} describes the latest developments in understanding the structure of extended operators in SCFTs, including their symmetry structure. 
While the first three sections focus on theories with eight or more supercharges, section \ref{sec:minSusy} instead summarizes the state of our understanding of theories with less supersymmetry.

Given length limitations, this white paper inevitably leaves out numerous 
results which connect to the core themes discussed below. 
We apologize in advance for these omissions.

\section{Classification} \label{sec:classification}

In this section we summarize classification efforts on SCFTs in various spacetime dimensions. We divide our discussion into the case of 
$D > 4$ SCFTs and $D \leq 4$ SCFTs. For $D > 4$, all Lagrangian interactions are irrelevant at weak coupling.

\section*{$D > 4$ SCFTs}

As mentioned, SCFTs can exist only for $D\leq 6$ \cite{Nahm:1977tg}.  For $D=6$ the classification of \cite{Nahm:1977tg} allows ${\cal N}=(n,0)$, corresponding to $N_Q=8n$ real supercharges of the same chirality. For all $D=3,4,6$ the infinite families admitted by \cite{Nahm:1977tg} only admit sensible SCFTs for $N_Q\leq 16$ supercharges ~\cite{Cordova:2016emh}, so only ${\cal N}=(1,0)$ and ${\cal N}=(2,0)$ are possible 6D SCFTs.  For $D = 5$, there is a unique superconformal algebra, and all such SCFTs have $\mathcal{N} = 1$ supersymmetry (i.e., eight real supercharges).  Again, all Lagrangian interactions for $D>4$ are irrelevant, so IR free, and the original arguments for the existence of non-trivial, interacting SCFTs in $D=5$~\cite{Seiberg:1996bd} and $D=6$~\cite{Seiberg:1996qx} dimensions included earlier hints from string theory brane constructions \cite{Witten:1995zh,Strominger:1995ac}. 

The SCFTs have a moduli space of supersymmetric vacua where the theory is weakly coupled.  Moduli spaces of SCFTs with $N_Q=8$ supercharges are similiar across dimensions for all $3\leq D\leq 6$: there can be a Higgs branch, and a Coulomb branch, and in 6D the Coulomb branch is actually a tensor-branch, associated with a tensor multiplet rather than a vector multiplet.  These branches touch at the origin, where there can be an interacting SCFT. 
%there can be a Coulomb branch where the dilaton is in a vector multiplet for $D\leq 5$, and in a tensor multiplet for $D=6$, and where the $SU(2)_R$ is the superconformal algebra.  The other possible branch is the Higgs branch, where $SU(2)_R$ is spontaneously broken and its NG bosons and the dilaton are in a hypermultiplet. 
In constructions based on string compactification, one considers a limit where gravity is switched off, and the background can be taken to be non-compact, with the field theory degrees of freedom localized at the singularities of the non-compact geometry.  The string constructions for generic, non-singular configurations give the SCFT at a generic point on the moduli space, where the theory is IR free.  To get the interacting SCFT, one must tune the string constructions to  singular configurations, corresponding to tuning the moduli fields to sit at the origin.

In $D=6$, there are supersymmetry protected (BPS) strings whose tension is proportional to the vev of the scalar parameterizing the tensor branch (the $D=6$ analog BPS $W$-bosons on the Coulomb branch), leading to ``tensionless strings'' as in \cite{Witten:1995zh, Strominger:1995ac}.  It was argued in ~\cite{Seiberg:1996qx} that these apparent tensionless strings can be understood (in the limit 
where gravity decouples) in terms of conventional QFT, associated with interacting SCFTs.\footnote{Let us note that the strings appearing at non-singular points of the moduli space are not the fundamental strings of superstring theory. Rather, they are effective strings more akin to the QCD effective string.} 
There are by now many constructions of 6D SCFTs, which include both early and more recent efforts, including references \cite{Witten:1995zh, Strominger:1995ac, Ganor:1996mu, Seiberg:1996vs, Witten:1996qb, Morrison:1996pp, Witten:1996qz, Bershadsky:1996nu, Brunner:1997gf, Blum:1997mm, Blum:1997fw, Aspinwall:1997ye, Intriligator:1997dh, Hanany:1997gh, Heckman:2013pva, DelZotto:2014hpa, Heckman:2015bfa}. An important comment here is that while some 6D SCFTs have a gauge theory description on the tensor branch, some do not, and in any case, all are defined by taking various singular limits in brane / geometric constructions.

The original examples of 5D interacting SCFTs were constructed by via $g^{-2}\to 0$ strong coupling limits of  gauge theories~\cite{Seiberg:1996bd}, so the SCFTs have a relevant deformation, $\Delta {\cal L} \sim g^{-2}F^{\mu \nu}F_{\mu \nu}+\dots$ which drive RG flows to the IR-free Lagrangian gauge theories; the IR-free gauge theories UV complete to the SCFTs.  The $g^{-2}$ relevant deformation of the SCFT is associated with a global symmetry (as are all supersymmetry-preserving, relevant deformations of 5D SCFTs), the instanton number $U(1)_I$ symmetry.   As in the case of 6D SCFTs, there need not exist a gauge theory phase in the moduli space, and the SCFTs can be obtained by taking singular limits of brane and geometric constructions. See \cite{Seiberg:1996bd, Katz:1996xe, Witten:1996qb, Morrison:1996xf, Douglas:1996xp, Intriligator:1997pq, Aharony:1997ju, Aharony:1997bh, Diaconescu:1998cn, Bergman:2012kr, Bergman:2013koa, DelZotto:2017pti, Jefferson:2018irk,Closset:2018bjz, Bhardwaj:2018yhy, Bhardwaj:2018vuu, Bhardwaj:2019fzv, Apruzzi:2019vpe, Apruzzi:2019opn, Apruzzi:2019enx, Tian:2021cif, DelZotto:2022fnw} for additional constructions of 5D SCFTs. 

\subsection{6D SCFTs}

%In six dimensions, the best evidence for the existence of 6D SCFTs comes from singular limits of string constructions in which all length scales are taken to either zero or infinity (the conformal case). As such, it is not surprising that the classification efforts all involve the combination of singular geometries and brane configurations.  A property of all of these constructions is that at a generic point of the moduli space, one finds effective strings which have a finite tension. At the singular locus where an SCFT exists, the tension of these effective strings passes to zero. 

All known 6D SCFTs have a tensor branch moduli space of vacua, parametrized by the expectation value of a real scalar in the tensor multiplet, which contains a 2-form gauge field.    There are BPS strings on this moduli space, and the SCFT is at the origin, where it seems that the tension of the effective strings goes to zero -- such apparently tensionless string limits are actually interacting SCFTs~\cite{Seiberg:1996qx}.  Perhaps the best known example of a 6D SCFT is the A-type $\cN = (2,0)$ theory, as given by the worldvolume theory of coincident M5-branes \cite{Strominger:1995ac}. In this case, the effective strings come from M2-branes which are stretched between M5-branes. One reason for the name ``A-type'' is that upon compactification on a $T^2$, this theory descends to $\cN = 4$ super Yang-Mills theory with gauge symmetry algebra $\mathfrak{su}(N)$ for $N$ M5-branes. There are also D- and E-type $\cN = (2,0)$ theories, and these can be obtained from considering type IIB string theory on an orbifold which preserves 16 supercharges.  The ADE classification of such orbifold singularities fits with an ADE classification of $\cN = (2,0)$ theories, see e.g.~\cite{Henningson:2004dh,Cordova:2015vwa}.\footnote{Compactification on a $T^2$ again gives rise to the corresponding $\cN = 4$ theory with respective D- and E-type gauge symmetry algebra. Non-simply laced algebras can also be obtained by twisting by an outer automorphism of the symmetry algebra. Likewise, the global structure of the gauge group (rather than just the Lie algebra) is dictated by restricting the spectrum of extended objects compactified on the $T^2$. See, e.g., \cite{Tachikawa:2013hya} and references therein for further discussion on this point.}

There are many more constructions of 6D $\cN = (1,0)$ SCFTs, as cited above.  A systematic approach to a possible classification of $\cN = (1,0)$ theories was undertaken in \cite{Heckman:2013pva, Heckman:2015bfa, Bhardwaj:2015xxa}. The main tool in this approach is to seek out the most general singular string backgrounds which could produce the requisite vanishing tension BPS strings.  A powerful method for accomplishing this is via F-theory \cite{Vafa:1996xn, Morrison:1996na, Morrison:1996pp}. There are many reviews of F-theory, see, e.g., \cite{Heckman:2010bq, Weigand:2018rez}. For our purposes, the main point is that the 10D spacetime of type IIB strings is supplemented by the profile of the IIB axio-dilaton, and this is captured by a $T^2$ fibration over the 10D spacetime. To make a 6D $\cN = (1,0)$ SCFT, one seeks out ``elliptically fibered Calabi-Yau threefolds'', namely a space $B$ of complex dimension two (i.e. real dimension four) and a $T^2$ which sits over each point of $B$ such that the total space is Calabi-Yau, a structure which is essentially the string theory analog of a Seiberg-Witten geometry \cite{Seiberg:1994rs, Seiberg:1994aj}.  In F-theory realizations of 6D SCFTs, the tensionless strings arise from D3-branes wrapped on collapsing two-cycles in the two complex-dimensional base $B$ of the F-theory threefold $X \rightarrow B$. Earlier studies of related F-theory backgrounds include \cite{Bershadsky:1996nu, Aspinwall:1997ye, Morrison:2012np}. The classification of type II and M-theory $AdS_7$ vacua was carried out in references \cite{Apruzzi:2013yva, Apruzzi:2015wna, Gaiotto:2014lca, Apruzzi:2017nck}.  A full classification of such F-theory geometries was accomplished in \cite{Heckman:2015bfa}.  It was subsequently realized that the same geometry can sometimes describe a ``frozen phase'' of F-theory, namely, the geometry could be related to different physical 6D SCFTs, requiring additional data to specify the SCFT \cite{Tachikawa:2015wka, Bhardwaj:2015oru, Bhardwaj:2018jgp, Bhardwaj:2019hhd}. For a more comprehensive review of 6D SCFTs, see reference \cite{Heckman:2018jxk}.

Moving to a point in moduli space where all effective strings have finite tension is known as the ``tensor branch'' of the theory (because it is achieved by giving vevs to the real scalars of the 6D supermultiplet known as the tensor multiplet).  Remarkably, the tensor branch of all such 6D SCFTs resembles a generalized quiver gauge theory with a single one-dimensional spine of gauge groups (and accompanying tensor multiplets) joined together by links known as ``conformal matter'' \cite{DelZotto:2014hpa, Heckman:2014qba}.  Conformal matter with flavor symmetry algebra $\mathfrak{g} \oplus \mathfrak{g}$ with $\mathfrak{g}$ a Lie algebra of ADE type is also realized by M5-branes probing the ADE singularity $\C^2 / \Gamma_{ADE}$. One important 
use for this quiver picture is to extract the structure of zero-form global 
symmetries \cite{Bertolini:2015bwa, Heckman:2016ssk, Morrison:2016djb, Lee:2018ihr, Apruzzi:2020eqi}, as well as 
their anomalies.  The explicit algorithm for computing these anomalies was established in references \cite{Harvey:1998bx, Intriligator:2014eaa, Ohmori:2014pca, Ohmori:2014kda, DelZotto:2014fia, Heckman:2015ola, Cordova:2020tij, Apruzzi:2020eqi}. 
Another direct use for this quiver picture is in the study of the worldvolume of effective strings of the theory, as in \cite{Haghighat:2013gba, Haghighat:2013tka, Haghighat:2014pva, Haghighat:2014vxa, Gadde:2015tra, Apruzzi:2016iac, Kim:2016foj, Shimizu:2016lbw, DelZotto:2016pvm, Apruzzi:2016nfr, DelZotto:2018tcj}.

In fact, all of the 6D SCFTs realized via F-theory can be obtained via a process of fission (Higgs branch and tensor branch deformations) and fusion (gauging a simple gauge group and adding a tensor multiplet), starting from a handful of progenitor theories based on M5-branes probing a Horava-Witten $E_8$ wall wrapping an ADE singularity \cite{Heckman:2018pqx}. This same analysis also serves to significantly constrain possible renormalization group flows. Supersymmetric deformations of 6D SCFTs are all triggered by operator vevs and are specified as Higgs branch deformations when the R-symmetry is broken, and tensor branch deformations when the R-symmetry is preserved \cite{Louis:2015mka, Cordova:2016xhm}. This is reflected in the geometry as complex structure deformations and Kahler deformations of the Calabi-Yau. Moreover, Higgs branch deformations are essentially classified by the hierarchical algebraic data associated with flavor symmetry breaking patterns \cite{Heckman:2015ola, Heckman:2016ssk, Bourget:2019aer, Apruzzi:2020eqi, Bourget:2022ehw}. Outstanding open problems in this direction include obtaining a full classification of possible renormalization group flows triggered by deformations of a fixed point, and using the quiver-like structure of these theories, (especially in the limit of a large number of gauge groups) to obtain additional data on the operator content of these theories (see, e.g., \cite{Nunez:2018ags, Filippas:2019puw, Bergman:2020bvi, Baume:2020ure, Heckman:2020otd}). Another important goal in this direction is to completely establish a six-dimensional analog of the $a$-theorem for supersymmetric flows. Some of the differences compared with the 4D proof based on dilaton scattering \cite{Komargodski:2011vj} were noted in \cite{Elvang:2012st} and were further elaborated on in the supersymmetric setting in \cite{Heckman:2021nwg}. An analytic proof for the special case of moduli space flows 
was given in \cite{Maxfield:2012aw, Cordova:2015vwa} for $\mathcal{N} = (2,0)$ theories, and for tensor branch flows of $\mathcal{N} = (1,0)$ theories was established in \cite{Cordova:2015fha}. A geometric study of more general flows was presented in \cite{Heckman:2015ola}, which served as a starting point for a ``proof by brute force'' of $a/c$-theorems for tensor 
branch and Higgs branch flows \cite{Heckman:2015axa}. It remains an outstanding open problem to give an analytic 
proof of the $a$-theorem for supersymmetric Higgs branch as well as mixed branch flows. This may also shed light 
on how to establish the 6D $a$-theorem without the assumption of supersymmetry.

\subsection{5D SCFTs}

Much recent progress has also been made in the classification of 5D SCFTs.  The early examples of these theories \cite{Seiberg:1996bd, Morrison:1996xf, Intriligator:1997pq} have since been generalized to a number of complementary constructions involving branes at singularities, and singular limits of string compactification geometries.  Again, the operating method for realizing a 5D SCFT is to start with an effective field theory with a moduli space of vacua which also has finite tension strings.  Moving to a point in moduli space where these strings have vanishing tension then yields the requisite SCFT. The most flexible known framework which encompasses all known examples is based on M-theory on non-compact singular Calabi-Yau threefolds. Such backgrounds preserve eight real supercharges. To get a 5D SCFT, one requires a holomorphic divisor (i.e., a real four-dimensional subspace) which collapses to a point.  An M5-brane wrapped on such a divisor generates an effective string in the 5D theory, and its tension goes to zero when the divisor collapses to zero size.  A priori, this is a different point in moduli space from where particles become massless. Though not necessary to realize a 5D SCFT, many have a gauge theory phase, and a corresponding $U(1)_I$ global symmetry which assigns an instanton number to the codimension-four gauge instanton configurations \cite{Lambert:2014jna,Tachikawa:2015mha} (its one-form current is dual to the second Chern class of the gauge fields). This provides an important constraint on the dynamics of effective strings in such models.

It is conjectured that all 5D SCFTs arise from collapsing configurations of divisors in Calabi-Yau threefolds, possibly accompanied by a quotient by a symmetry (i.e., an automorphism) of the Calabi-Yau geometry.  This is closely related to the open mathematical question of classifying canonical singularities.  In the case of a small number of collapsing surfaces, there is a complete classification of such theories \cite{Jefferson:2017ahm, Jefferson:2018irk}.  For higher rank theories, much of the analysis has centered on the special case obtained from compactification of 6D SCFTs on a circle, possibly accompanied by an automorphism twist \cite{DelZotto:2017pti, Apruzzi:2018nre, Apruzzi:2019vpe, Apruzzi:2019opn, Apruzzi:2019enx, Apruzzi:2019kgb}.  In geometric terms, this occurs because 6D SCFTs arise from compactification of a canonical singularity of an elliptically fibered Calabi-Yau threefold, and so a further circle compactification, accompanied by Wilson lines and an automorphism twist, can flow to a 5D SCFT, as obtained from M-theory compactified on a Calabi-Yau threefold with a canonical singularity see e.g. \cite{Morrison:1996xf,Intriligator:1997pq,Xie:2017pfl,Closset:2020scj,Closset:2020afy,Closset:2021lwy}.

One of the major open questions in this direction is to find all 5D SCFTs obtained from canonical singularities for non-compact Calabi-Yau threefolds. The analogous question for theories with a dual $AdS_6$ supergravity description has largely been answered implicitly in a series of papers \cite{Brandhuber:1999np,Passias:2012vp,Lozano:2012au,Apruzzi:2014qva, DHoker:2016ujz, DHoker:2016ysh, Gutowski:2017edr, DHoker:2017mds, Gutperle:2018vdd,Uhlemann:2019ypp,Legramandi:2021uds}.  The classification of all Calabi-Yau geometries which can produce a 5D SCFT is more challenging than the classification of 6D SCFTs because there is no general numerical criterion for determining when configurations of collapsing divisors will produce a 5D SCFT, and so examples are handled on a case by case basis. It has been conjectured, however, that all 5D SCFTs arise from circle compactification (with twists) from a 6D SCFT \cite{DelZotto:2017pti}.  If this were true, the question would reduce to determining all possible moduli space flows in that more limited setting. Examples which resist an obvious embedding in an elliptically fibered Calabi-Yau threefold include orbifolds of the form $\C^3 / \Gamma$ for $\Gamma$ a finite subgroup of $SU(3)$ \cite{DelZotto:2017pti, Xie:2017pfl, Tian:2021cif, DelZotto:2022fnw}. That said, these examples may be connected to other theories via a flow in moduli space.

%Some relevant references: \cite{Intriligator:1997pq,Benini:2009gi,Bergman:2015dpa,Bergman:2013aca,Bhardwaj:2020gyu,Bhardwaj:2020ruf,Bhar%dwaj:2020avz,Bhardwaj:2019jtr,Bhardwaj:2018vuu,VanBeest:2020kxw,vanBeest:2020kou,Apruzzi:2019opn,Apruzzi:2019vpe,Apruzzi:2019enx,Apruzzi%:2019kgb,Bhardwaj:2018yhy,Jefferson:2018irk}

\section*{$D \leq 4$ SCFTs}

For $D\leq 4$ the gauge coupling is either marginal or relevant. Thus, gauge theories provide an extremely useful view into the space of interacting SCFTs in four dimensions or lower. Despite that, the landscape of SCFTs in $D\leq 4$ remains overwhelmingly populated by theories which do not have any Lagrangian description, though some can be obtained as IR fixed points of renormalization group flows, including some cases where supersymmetry is broken along the flow but restored or enhanced in the IR \cite{Maruyoshi:2016tqk, Agarwal:2016pjo, Agarwal:2017roi, Giacomelli:2017ckh, Benvenuti:2017bpg}.

\subsection{4D $\cN\geq 2$ SCFTs}

For 4D SCFTs with $\mathcal{N} = 2$ supersymmetry which have a gauge theory description, 
the gauge coupling is exactly marginal \cite{Green:2010da}.  Associated with the Lagrangian descriptions of these $\cN\ge2$ gauge theories \cite{Bhardwaj:2013qia} are a set of techniques based on supersymmetric localization, special geometric structures on conformal manifolds, and large $N$ and large R-charge limits, which permit the calculation of some more detailed observables --- in particular, correlators of non-chiral Coulomb branch operators in these SCFTs \cite{Lee:1998bxa, Papadodimas:2009eu, Baggio:2014ioa, Baggio:2014sna, Baggio:2015vxa, Gerchkovitz:2016gxx, Rodriguez-Gomez:2016ijh, Baggio:2016skg, Bourget:2018obm, Seiberg:2018ntt, Beccaria:2018xxl, Grassi:2019txd, Beccaria:2020hgy, Beccaria:2021hvt, Hellerman:2017veg, Hellerman:2017sur, Hellerman:2018xpi, Beccaria:2020azj, Hellerman:2021duh}.  In some cases these techniques have been extended to non-Lagrangian SCFTs via renormalization group flows from gauge theories \cite{Hellerman:2017veg, Grassi:2019txd, Bissi:2021rei}. Nevertheless, with the exception of the simplest set of theories, the so-called rank-1 theories \cite{Argyres:2015gha, Argyres:2016xmc, Argyres:2016xua, Argyres:2015ffa}, which have been completely classified, and partial progress for rank-2 theories \cite{Martone:2021ixp,Kaidi:2021tgr}, the classification of $\cN=2$ theories in four dimensions remains an open problem.

Dimensional reduction works in a remarkably effective way to engineer superconformal fixed points in four dimensions. The \emph{class-$\mathcal{S}$ theories}, i.e., those which can be engineered by compactifying $(2,0)$ 6D theories on a Riemann surface in the presence of codimension-two defects \cite{Chacaltana:2010ks, Chacaltana:2011ze, Chacaltana:2012zy, Chacaltana:2012ch, Chacaltana:2013oka, Chacaltana:2014jba, Chacaltana:2014nya, Chacaltana:2015bna, Chacaltana:2016shw, Chacaltana:2017boe, Chacaltana:2018vhp, Distler:2021cwz, Xie:2012jd, Beem:2020pry, Xie:2012hs, Wang:2015mra, Wang:2018gvb, Tachikawa:2018rgw} fill a vast landscape of SCFTs.  More recently, a thorough understanding of the wordvolume theory of D3 branes in F-theory probing both an $S$-fold \cite{Garcia-Etxebarria:2015wns, Aharony:2016kai} --- a generalization of an orientifold --- and an exceptional 7-brane singularity, the so-called $\cN=2$ S-folds \cite{Aharony:2016kai, Apruzzi:2020pmv, Giacomelli:2020jel, Heckman:2020svr, Giacomelli:2020gee, Bourget:2020mez} has been used to move beyond class $\mathcal{S}$.\footnote{The ``S'' of an S-fold has no relation to the ``$\mathcal{S}$'' of class $\mathcal{S}$ theories (which refers to $\mathcal{S}$ix dimensional).}  This construction also realizes theories which had been predicted by the systematic analysis of rank-1 theories and which can be instead realized in the more general construction of twisted compactification of 6D $(1,0)$ theories \cite{Ohmori:2018ona}. The methods which worked to compile the classification in rank-1 are of limited use for higher ranks. Thus for general answers a variety of approaches have been employed: $(a)$ leveraging singularity theory to perform a systematic analysis of $\cN=2$ SCFTs which can be engineered in type IIB on a Calabi-Yau threefold \cite{Xie:2015rpa, Chen:2016bzh, Wang:2016yha, Chen:2017wkw, Xie:2021hxd}, $(b)$ systematic study of the constraint of special K\"ahler geometry \cite{Freed:1997dp} constraining the set of allowed Coulomb branches  \cite{Cecotti:2021ouq, Martone:2020nsy, Argyres:2020wmq, Martone:2021ixp, Martone:2021drm, Argyres:2020nrr, Caorsi:2018zsq, Argyres:2018zay, Caorsi:2018ahl, Caorsi:2019vex, Kaidi:2021tgr}, and $(c)$ systematic understanding of Higgs branches of $N_Q=8$ supercharge SCFTs as the Coulomb branch of 3D $\cN=4$ SCFT magnetic quivers (as 3D mirror symmetry \cite{Intriligator:1996ex}) for Higgs branches of $D>3$ SCFTs\cite{Cabrera:2018jxt,Bourget:2019rtl, Cabrera:2018ldc, Bourget:2019aer, Bourget:2020asf, Baume:2020ure, Closset:2020scj, Closset:2020afy,Closset:2021lwy, Bourget:2021csg, Bourget:2021jwo}. These approaches are helpful given the limited effectiveness of bootstrap methods in this context \cite{Beem:2014zpa}.

Much more progress can be made studying theories in 4D with $\cN\geq3$ supersymmetry \cite{Aharony:2013hda, Aharony:2015oyb, Garcia-Etxebarria:2015wns, Bourton:2018jwb, Aharony:2016kai, Garcia-Etxebarria:2016erx, Garcia-Etxebarria:2017ffg}.  In this case it can be shown that the K\"ahler metric is necessarily flat \cite{Cordova:2016xhm, Argyres:2019yyb} and thus isotrivial\footnote{On the 
Coulomb branch of an $\mathcal{N} = 2$ theory, the metric on the Coulomb branch is generated from $\mathrm{Im} \tau_{ij}$, 
and the condition that the metric is isotrivial is just the statement that $\tau_{ij}$ is locally constant.} \cite{Cecotti:2021ouq}. Further restricting to theories with freely generated Coulomb branch chiral ring \cite{Argyres:2018wxu, Bourget:2018ond}, the moduli space of $\cN\geq 3$ theories in 4D are realized as orbifolds by crystallographic complex reflection groups \cite{Argyres:2019ngz, Caorsi:2018zsq, Bonetti:2018fqz}.  Many of the consistent low-energy solutions, particularly those which are associated with exceptional crystallographic complex reflection groups, have not been realized in string theory and thus remain conjectural.  Furthermore, an altogether new set of 4D SCFTs might arise upon lifting the assumption that the Coulomb branch chiral ring is freely generated; this would result in complex singularities in the moduli space of vacua \cite{Argyres:2017tmj, Argyres:2018wxu}.

Given the availability of techniques allowing classification schemes for SCFTs in 4D which do not rely on string theory, it is intriguing to compare these results with those obtained from compactification of the aforementioned 5D and 6D theories. The analysis is complicated by the variety of compactifications which are allowed \cite{Zafrir:2016wkk,Ohmori:2018ona} and there are currently only limited results in this direction. Preliminary evidence shows that many 4D SCFTs can indeed be obtained from SCFTs in higher dimensions \cite{Martone:2021drm}, though many challenges remain, e.g. see the classification of~\cite{Bhardwaj:2013qia} of 4D Lagrangian theories; many of the quivers there have no known stringy construction.\footnote{We thank Y. Tachikawa for this comment.} Understanding these challenging examples would be of great value in testing the possible completeness of string theory constructions of SCFTs.

\subsection{3D $\cN\geq 4$ SCFTs} 

By reducing 4D $\cN\geq2$ SCFTs on a circle, we obtain $\cN\geq4$ theories in 3D.  For 3D $\cN=4$ theories, 
the Coulomb branch has a hyperkahler structure which is similar to the one present on the Higgs branch. 
Indeed, this is compatible with 3D mirror symmetry duality \cite{Intriligator:1996ex}, which exchanges the  Higgs branch of theory $\cT^A$ with the Coulomb branch of $\cT^B$ --- the \emph{mirror dual} of $\cT^A$ --- and vice versa.  Recently there has been a flurry of activity which has improved our general understanding of mirror duals of Argyres-Douglas theories \cite{Beratto:2020wmn, Giacomelli:2020ryy, Carta:2021whq, Xie:2021ewm, Carta:2021dyx, Dey:2021rxw} as well as the identification of new mirror dualities altogether \cite{Nedelin:2017nsb, Aprile:2018oau, Dey:2020hfe}.  Further insights into mirror symmetry can also be obtained \cite{Dedushenko:2017avn, Fan:2019jii} by exploiting the fact that 3D $\cN=4$ theories have a topological sector \cite{Chester:2014mea, Beem:2016cbd}, which descends from the vertex operator algebra (VOA) associated to $\cN=2$ 4D SCFTs \cite{Dedushenko:2019mzv, Dedushenko:2019mnd} (see below for additional details on the appearance of VOAs).

Despite all this recent progress, a systematic charting of $\cN=4$ 3D theories remains a vast challenge. The situation is more manageable upon restricting to $\cN\geq5$, see ~\cite{Hosomichi:2008jb} for $\cN=5$ examples, for $\cN\geq 6$ see \cite{Aharony:2008ug, Aharony:2008gk, Schwarz:2004yj, Bagger:2007jr, Gustavsson:2007vu,Chu:2009gi,Chu:2010fk,Gran:2012mg} and, for $\cN =8$ maximally supersymmetric theories in 3D, see \cite{Gran:2008qx,Nilsson:2013fya,Agmon:2017lga, Bergman:2020ifi, Binder:2021cif, Alday:2021ymb}.
In particular there has been an intriguing proposal for a classification scheme of $\cN\geq6$ theories based on real and complex reflection groups \cite{Tachikawa:2019dvq}.  If true, this predicts the existence of two new $\cN=8$ SCFTs. In three dimensions, $\cN=7$ supersymmetry necessarily implies $\cN=8$ supersymmetry \cite{Bashkirov:2011fr,Cordova:2016emh} and SCFTs with $N_Q>16$ can exist but are necessarily free \cite{Cordova:2016emh}.

\section{SCFTs and new mathematics}\label{sec:newMath}

SCFTs have also been used as a powerful tool to glean insights in a variety of separate fields in mathematics.  We will here just mention a few examples, leaving a fuller discussion  on the intersection between quantum field theory and mathematics to \cite{PhysMat}.

\subsection{Geometry and singularities} 

The close interplay between singular Calabi-Yau geometries and the resulting physical theories has been fruitful in both directions. In particular, physical considerations can predict new mathematical structures. The expectation that physical theories can be connected under renormalization group flows is manifested in the geometry as a hierarchical stratification of singular geometries according to a partially ordered set.  As explicit examples, in 6D SCFTs, a number of flavor symmetry breaking patterns are captured by nilpotent orbits of elements in the flavor symmetry algebra \cite{Cremonesi:2015bld, Heckman:2016ssk, Hanany:2016gbz, Mekareeya:2016yal, Heckman:2018pqx, Apruzzi:2018xkw, Bourget:2019aer, Hassler:2019eso, Baume:2021qho}. These in turn admit a partial ordering, which directly translates to smoothing deformations of the associated geometry. By the Jacobson-Morozov theorem, each nilpotent orbit defines a Lie algebra homomorphism $\mathfrak{sl}(2) \rightarrow \mathfrak{g}_{\mathrm{flav}}$, in the obvious notation. Another application of related mathematical structures is the development of a geometric classification scheme for finite group homomorphisms from finite subgroups of $SU(2)$ to the Lie group $E_8$ \cite{Heckman:2015bfa, Frey:2018vpw}, which was even used to correct a few typos in the 
original list of examples presented in reference \cite{FREY}! The common bridge connecting these seemingly different structures is via the 6D SCFT of M5-branes probing an ADE singularity wrapped by a Horava-Witten nine-brane. An open problem here is to use the 
hierarchy of RG flows in physics to develop a partial ordering of such homomorphisms.

The same sort of geometric partial ordering also appears in Higgs branch deformations of 5D SCFTs \cite{Apruzzi:2019opn, Apruzzi:2019enx}.  In the case of 6D SCFTs, the classification of canonical singularities for elliptically fibered Calabi-Yau threefolds was accomplished using physical methods in \cite{Heckman:2015bfa}.  This was also recently joined by a mathematical classification scheme in \cite{SvaldiDiCerbo}. The case of $D = 5$ SCFTs involves the mathematically far more challenging issue of classifying canonical singularities of non-compact Calabi-Yau threefolds, i.e., relaxing the condition that there is an elliptic fibration.  See e.g.~\cite{Closset:2020scj,Closset:2021lwy} for studies of canonical singularities related to 4D and 5D SCFTs.

In lower-dimensional SCFTs, the possible geometric singularities which can be realized in the ``internal'' compactification directions can become significantly more intricate.  For example, for Calabi-Yau spaces of complex dimension four and above, it is possible to have singularities which do not admit a crepant resolution.  These figure prominently in the physics of S-folds (see, e.g., \cite{Garcia-Etxebarria:2015wns, Garcia-Etxebarria:2016erx, Aharony:2015oyb, Nishinaka:2016hbw, Aharony:2016kai, Imamura:2016abe, Imamura:2016udl, Agarwal:2016rvx, Cordova:2016emh, Lemos:2016xke, Arras:2016evy, Lawrie:2016axq, vanMuiden:2017qsh, Amariti:2017cyd, Bourton:2018jwb, Assel:2018vtq, Tachikawa:2018njr, Ferrara:2018iko, Bonetti:2018fqz, Garozzo:2018kra, Arai:2018utu, Garozzo:2019hbf, Arai:2019xmp, Garozzo:2019ejm, Amariti:2020lua, Zafrir:2020epd, Heckman:2020svr, Giacomelli:2020jel, Giacomelli:2020gee, Bourget:2020mez, Kimura:2020hgw}), which have already led to the discovery of many new 4D $\cN \leq 3$ SCFTs.  What is currently unclear is how severe a singularity can be admitted whilst still retaining a physical SCFT interpretation. For recent discussion on the physical interpretation of terminal singularities, see, e.g., \cite{Grassi:2018rva,Closset:2020scj}.

\subsection{Vertex operator algebras}

To any four-dimensional $\cN=2$ SCFT $\cT$ one can canonically associate a two-dimensional non-unitary vertex operator algebra (VOA) $\chi[\cT]$ \cite{Beem:2013sza} which arises as a cohomological reduction of the full OPE algebra of the four-dimensional theory, or equivalently, by introducing a certain $\Omega$ background that deforms the holomorphic-topological twist of the theory \cite{Jeong:2019pzg, Oh:2019bgz}. $\chi[\cT]$ carries many features of the 4D avatar: its central charge is $c_{2D}=-12c_{4D}$ and the 4D flavor symmetry gets enhanced to an affine Lie algebra and $k_{2D}=-k_{4D}/2$.  Less directly $\chi[\cT]$ is constrained by 4D unitarity \cite{Liendo:2015ofa, Lemos:2015orc, Beem:2018duj} but it remains an open problem how to completely characterize it.  Since the early days, it has been noticed that the structure of $\chi[\cT]$ is deeply connected with the physics of the Higgs branch which conjecturally \cite{Beem:2017ooy} arises as the associated variety \cite{Arakawa:2010ni} to the VOA.  
This conjecture, which has been shown \cite{Dedushenko:2019mzv} to imply a previous conjecture of Arakawa \cite[Conjecture 1]{Arakawa:2015}, carries deep implications: the VOAs which arise from the cohomological reduction of $\cN=2$ SCFTs would then be of a special type known as ``quasi-lisse" \cite{Arakawa:2016hkg}, a property which ensures that their vacuum characters satisfy a linear modular differential equation \cite{Buican:2015ina, Buican:2015tda, Cordova:2015nma}.  The representation theory of $\chi[\cT]$, which is seldomly rational, can be complicated and it remains an open problem to understand which characters participate in the modular property of the vacuum character; these and related issues have been recently investigated in \cite{Kang:2021lic, Pan:2021ulr, Pan:2021mrw, Beem:2021zvt}.  In recent years a varieties of techniques have been employed to compute $\chi[\cT]$ in a large set of examples \cite{Beem:2014rza, Lemos:2014lua, Xie:2019vzr, Xie:2019zlb, Song:2017oew, Xie:2016evu, Buican:2016arp, Buican:2015tda, Buican:2015ina, Fluder:2017oxm, Song:2016yfd, Song:2015wta, Dedushenko:2019mnd, Dedushenko:2019yiw, Agarwal:2021oyl} and it is worthwhile noticing that surprisingly often, in the Argyres-Douglas case, $\chi[\cT]$ is an affine Kac-Moody at boundary admissible level \cite{Song:2017oew, Xie:2016evu, Xie:2019yds, kac2016remark}. 

The connection between VOA and Higgs branch physics might be even deeper than initially thought.  In fact, inspired by \cite{Adamovic:2004zi}, it has been recently shown that the low-energy effective theory on the Higgs branch provides a way to build free field realizations of $\chi[\cT]$ \cite{Bonetti:2018fqz, Beem:2019tfp, Beem:2019snk,  Beem:2021jnm}.  This perspective might provide an interesting framework to leverage VOA constraints to classify allowed 4D SCFTs.  The VOA is moreover connected~\cite{Fluder:2017oxm} to the Lens index. Finally, tantalizing connections between chiral algebras and Coulomb branches have been noticed in a variety of examples \cite{Cordova:2015nma, Fredrickson:2017yka, Dedushenko:2018bpp}.  It remains an open question whether this is a general property of VOAs, which would then be even more constrained than initially expected.

\section{Extended operators and higher-form symmetries}\label{sec:extOp}

Extended operators are important observables of QFT.  For example, the spectrum of Wilson-'t Hooft line operators in 4D gauge theories is non-perturbative data which differentiates between otherwise perturbatively equivalent QFTs \cite{Aharony:2013hda}.   It is challenging to characterize the spectra of such operators, and compute their correlators, in general QFTs. See e.g. \cite{Kapustin:2006pk, Gukov:2006jk, Gukov:2007ck, Kapustin:2007wm, Gukov:2008sn, Gaiotto:2009fs, Tan:2009he, Tan:2009qq, Tan:2010dk, Bruzzo:2010fk, Kanno:2011fw} for aspects of topological extended operators. 
%A general theory of correlators of extended operators is daunting partly because the configuration space of extended operators is infinite-dimensional.  
Supersymmetry-protected (BPS) extended operators are better understood via 
techniques such as supersymmetric localization, and via their geometric realization as wrapped branes in string / M- / F-theory constructions \cite{Constable:2002xt, Buchbinder:2007ar, Gomis:2007fi, Pestun:2007rz, Drukker:2007dw, Drukker:2007yx, Drukker:2007qr, Koh:2008kt, Drukker:2008wr, Koh:2009cj, Pestun:2009nn, Giombi:2009ek, Giombi:2009ds, Alday:2009fs, Dimofte:2010tz, Passerini:2010pr, Gomis:2014eya, DelZotto:2015isa, Ashok:2017lko, Pan:2017zie, Ashok:2018zxp, Razamat:2018zel, Bianchi:2020hsz, Morrison:2020ool, Albertini:2020mdx}.

For CFTs, a key simplification comes from focusing on ``conformal extended operators" \cite{Kapustin:2005py}, i.e. those configurations which preserve a maximal subgroup of the conformal group.  These are the flat or straight extended operators and their conformal transforms; in euclidean space they have planar or spherical world volumes.  Our understanding of the constraints on the spectrum of extended operators coming from CFT unitarity and of the structure of the operator product algebra of conformal extended operators is rapidly developing \cite{Bootstrap,Billo:2016cpy, Gadde:2016fbj, Drukker:2017xrb, Giombi:2017cqn, Kim:2017sju, Fukuda:2017cup, Beccaria:2017rbe, Lauria:2017wav, Lemos:2017vnx, Kobayashi:2018okw, Guha:2018snh, Lauria:2018klo, DiPietro:2019hqe, Bianchi:2019umv, Lauria:2020emq, Herzog:2020bqw, Giombi:2021uae, Bianchi:2021snj}. Challenges, such as generalizing the notions of primary and descendant operators for extended operators, may become more tractable by specializing to BPS conformal operators in SCFTs; some work along this direction is \cite{Gomis:2009ir, Gaiotto:2009fs, Gaiotto:2012xa, Gadde:2013dda, Gaiotto:2013sma, Cordova:2016uwk, Gorsky:2017hro, Cordova:2017ohl, Cordova:2017mhb, Drukker:2017dgn, Bianchi:2018zpb, Nishinaka:2018zwq, Bianchi:2019sxz, Agmon:2020pde, Wang:2020xkc}

Generalized, $n$-form global symmetries \cite{Gaiotto:2014kfa} can act on $m$-dimensional extended operators for $m\ge n$. Thus local operators are charged only under ordinary ($0$-form) symmetries, line operators may be charged under both $0$- and $1$-form symmetries, etc.  $n$-form symmetry charges and transformations are captured by the insertion of $(d{-}n{-}1)$-dimensional extended operators in correlators.  These symmetry operators are topological operators, and their action is sensitive only to the topological linking of their world volumes with the world volumes of other operators in the correlator.  The generalized symmetry operators may themselves be charged under $n$-form symmetries, implying that different $n$-form symmetries may have an ``extended group" structure.  Moreover, $n$-form symmetries may be part of larger ``non-invertible symmetries" consisting of the algebra of all topological operators in a theory. The characterization of higher-form and non-invertible symmetries in various QFTs is an area of active research, see \cite{Topological}.  

Supersymmetry, and the moduli spaces of vacua, helps to determine these symmetry algebras and their action on the spectrum of operators. Superconformal symmetry provides strong additional restrictions.  Indeed, the representation theory of the superconformal algebra forbids higher-form conserved currents, so there can be no continuous higher-form symmetries in SCFTs \cite{Cordova:2016emh}; see also \cite{Lee:2021obi} for similar restrictions without assuming supersymmetry. There can be continuous higher symmetry in non-conformal supersymmetric theories, and its presence is an obstruction to conformal invariance anywhere along the RG flow unless the symmetry is accidental in the IR and explicitly broken in the UV \cite{Cordova:2018cvg,Cordova:2020tij}. SCFTs {\it can} have interesting {\it discrete} higher-form symmetries, and this has been an active area of recent research: recent work in six dimensions includes 
\cite{DelZotto:2015isa,Heckman:2017uxe,Closset:2020scj,DelZotto:2020sop,Apruzzi:2020zot, Bhardwaj:2020phs,Apruzzi:2021mlh}; in five dimensions includes   \cite{Morrison:2020ool,Bhardwaj:2020phs,Bhardwaj:2021zrt,Apruzzi:2021vcu, DelZotto:2022fnw}; in four dimensions includes \cite{DelZotto:2020esg,Albertini:2020mdx,Bhardwaj:2021pfz,Bhardwaj:2021wif,Bhardwaj:2021mzl}; and in three dimensions includes \cite{Kapustin:2014lwa, Benini:2017dus,Benini:2018reh, Hsin:2018vcg,Eckhard:2019jgg,Beratto:2021xmn,Cvetic:2021maf}.

%helps to constrain the symmetry algebras and In general it is difficult to compute these symmetry algebras and their action on the spectrum of operators in strongly-interacting QFTs.  This situation is ameliorated with the addition of superconformal symmetry.  
%In string/M theory constructions, susy-protected extended operators can be seen geometrically via branes wrapping compactification cycles. 

%topological extended op refs:
%\cite{Kapustin:2006pk, Gukov:2006jk, Gukov:2007ck, Kapustin:2007wm, Gukov:2008sn, Gaiotto:2009fs, %Tan:2009he, Tan:2009qq, Tan:2010dk, Bruzzo:2010fk, Kanno:2011fw}

\section{Theories with less supersymmetry ($8>N_Q\geq 2$)}\label{sec:minSusy}
Above we considered SCFTs with $N_Q=8$ supercharges. Here we mention aspects of SCFTs with $N_Q=4$ supercharges, namely 4D $\cN=1$ and 3D $\cN = 2$ theories, where the superconformal algebra contains a $U(1)_R$ symmetry. Many examples of 4D $\cN=1$ SCFTs were found starting in the mid 1990s, via singularities of the moduli space, dualities and 't Hooft anomaly matching, and a variety of other methods, see e.g.~\cite{Seiberg:1994bz,Intriligator:1994sm,Seiberg:1994pq,Kutasov:1995ve,Kutasov:1995np,Intriligator:1995ax,Intriligator:1995au,Chaichian:2000wr} and references therein. 
Additional checks of 4D $\cN=1$ dualities and the SCFT operator spectrum comes from superconformal indices, see e.g. \cite{Dolan:2008qi,Spiridonov:2009za,Closset:2017bse}. In 4D $\mathcal{N} =1$ finding the exact superconformal $U(1)_R$ can require $a$-maximization~\cite{Intriligator:2003jj}

%3D allows also for the possibility of supersymmetrized Chern-Simons terms. This makes the landscape of dualities in 3D $\cN=2$ quite rich see e.g.  \cite{deBoer:1997kr,Aharony:1997bx,Aharony:1997gp,Giveon:2008zn,Benini:2011mf,Intriligator:2013lca,Amariti:2020xqm} and there is an active effort in fully characterizing 3D $\cN=2$ dualities beyond theories with $SU(N)$ gauge group and/or with matter in representations fundamental/adjoint representations \cite{Okazaki:2021gkk,Fazzi:2018rkr,Amariti:2020xqm,Benvenuti:2018bav,Amariti:2018wht}. Relatedly, these dualities involve the non-trivial matching of electric and magnetic degrees of freedom on the two sides and it is particularly challenging to fully understand the properties of monopole operators \cite{Amariti:2015kha,Benini:2017dud,Amariti:2018gdc,Benvenuti:2016wet}. Another challenge is that in 3D there is only discrete anomalies, e.g. the $\Z_2$-valued parity anomaly \cite{Redlich:1983kn,Redlich:1983dv, Witten:2016cio} which makes it challenging to provide definitive evidence for these dualities. Fortunately, computing exact operator dimensions and other exact results localization techniques in 3D $\cN=2$ \cite{Kapustin:2009kz} yield highly non-trivial evidence for the many conjectured 3D dualities, see e.g.  \cite{Willett:2011gp,Closset:2018ghr,Closset:2019hyt,Jain:2019lqb}. 

Dimensional reduction from 6D SCFTs is an extremely useful tool in the construction and study of $N_Q=4$ SCFTs.   It is possible to obtain 4D $\cN=1$ SCFTs  from compactification of $\mathcal{N} = (2,0)$ and $\mathcal{N} = (1,0)$ 6D SCFTs. To obtain an $\cN=1$ SCFT in the former case, one has to turn on fluxes for the global symmetries \cite{Benini:2009mz, Bah:2012dg,Beem:2012yn,Agarwal:2015vla,Fazzi:2016eec,Nardoni:2016ffl}.  In the latter case instead it is possible to compactify on a generic Riemann surface \cite{Morrison:2016nrt, Kim:2017toz, Kim:2018lfo, Kim:2018bpg, Apruzzi:2018oge, Razamat:2019ukg, Hwang:2021xyw}, with additional boundary conditions for operators specified at marked points or ``punctures'' \cite{Gaiotto:2009we, Xie:2013gma, Bah:2013aha, Agarwal:2013uga, Agarwal:2014rua, Heckman:2016xdl, Hassler:2017arf}. Particularly tractable cases are theories obtained from compactification of the 6D E-string theory \cite{Nazzal:2018brc,Pasquetti:2019hxf,Razamat:2020bix} and the theories of class $\mathcal{S}_k$ obtained from a stack of M5-branes probing an $A_{k-1}$ singularity \cite{Gaiotto:2015usa, Coman:2015bqq, Franco:2015jna, Hanany:2015pfa, Razamat:2016dpl, Bah:2017gph, Bourton:2020rfo}.  The interesting pattern that arises in many of these examples is that the simplicity of the parent 6D theory translates to the simplicity of the 4D $\cN=1$ SCFTs with a variety of minimal $\cN=(1,0)$ theories having a simple quiver formulation \cite{Razamat:2018gro}. For many other such $\mathcal{N} = 1$ SCFTs, it remains an outstanding open problem to determine if the SCFTs can also be defined in terms of RG flows from 4D Lagrangian theories. See also \cite{Gadde:2013fma,Gadde:2015xta, Agarwal:2018ejn, Sabag:2020elc,Chen:2019njf,Zafrir:2018hkr} for other examples of 4D $\cN=1$ SCFTs and dualities obtained via compactification; for efforts to charting the possible 4D $\cN=1$ SCFTs and their connections via RG flows and dualities see e.g.~\cite{Maruyoshi:2018nod} where it is shown that even highly restricted UV starting points can lead to a remarkably rich landscape of SCFTs. See also~\cite{Agarwal:2020pol} for a classification of rich classes of SCFTs obtained from simple gauge group with a large $N$ limit with dense spectrum. 

An another recently discovered intriguing phenomenon is the existence of a 4D $\cN=1$ conformal manifold for theories with higher supersymmetry, even in the absence of exactly marginal deformations fully preserving the supersymmetry of the original theory \cite{Maruyoshi:2016aim, Razamat:2019vfd, Razamat:2020pra}. This observation has led to the discovery of new $\cN=1$ dualities \cite{Zafrir:2019hps, Razamat:2020gcc} as well as an $\cN=1$ Lagrangian for an inherently strongly coupled $\cN=3$ theory \cite{Zafrir:2020epd}.

Finding the exact $U(1)_R$ symmetry in 3D $\cN=2$ theories can require $Z$-extremization \cite{Jafferis:2010un,Closset:2012vg} (or $\tau _{RR}$ extremization \cite{Barnes:2005bm}, which applies for SCFTs in any $D$, see also \cite{Closset:2012ru,Nishioka:2013gza,Amariti:2015ybz,Amariti:2021cpk}).
In 3D there are also $\cN=3$ and $\cN=1$ supersymmetries with $N_Q=6$ and $N_Q=2$, respectively.\footnote{Localization with $\cN=3$ supersymmetry allows for more detailed calculations in the matching of scaling dimensions of operators of dual theories \cite{Kapustin:2010mh,Jafferis:2010un}, and aspects of this case have been explored in  \cite{Giveon:2008zn, Kubo:2021ecs}.  
In the 3D $\cN=1$ case, the analysis is more difficult as there is no continuous R-symmetry and the superpotential is real, not holomorphic, see e.g. \cite{Gaiotto:2018yjh,Benini:2018umh,Cvetic:2021maf} and references therein for aspects of $\cN=1$ theories.} 
Furthermore, compactifying to three dimensions allows for additional structure since the interactions are no longer constrained by asymptotic freedom bounds and there is the possibility of supersymmetrized Chern-Simons terms.  The landscape of SCFTs and dualities in 3D $\cN=2$, and inter-connections with compactified higher dimensional theories, is thus quite rich, see e.g.\   \cite{deBoer:1997kr, Aharony:1997bx, Aharony:1997gp, Giveon:2008zn, Benini:2011mf, Intriligator:2013lca, Amariti:2020xqm, Sacchi:2021wvg, Sacchi:2021afk}, and there is an active effort in fully characterizing 3D $\cN=2$ dualities beyond theories with $SU(N)$ gauge group and/or with matter in representations fundamental/adjoint representations \cite{Okazaki:2021gkk,Fazzi:2018rkr,Amariti:2020xqm,Benvenuti:2018bav,Amariti:2018wht,Giacomelli:2019blm}. 
Relatedly, these dualities involve the non-trivial matching of electric and magnetic degrees of freedom on the two sides and it is particularly challenging to fully understand the properties of monopole operators \cite{Amariti:2015kha,Benini:2017dud,Amariti:2018gdc,Benvenuti:2016wet}. 
Another challenge is that in 3D there are only discrete anomalies, e.g. the $\Z_2$-valued parity anomaly \cite{Redlich:1983kn,Redlich:1983dv, Witten:2016cio}, which makes it challenging to provide definitive evidence for these dualities. Fortunately, computing exact operator dimensions and other exact results localization techniques in 3D $\cN=2$ \cite{Kapustin:2009kz} yield highly non-trivial evidence for the many conjectured 3D dualities, see e.g.  \cite{Willett:2011gp, Closset:2018ghr, Closset:2019hyt, Jain:2019lqb}.

Finally, there is an intriguing connection between 3D $\cN=2$ and the characterization of three manifolds. This is done by exploiting the fact that compactifying a 6D (2,0) theory on a three manifold $\cM_3$ gives a 3D $\cN=2$ theory which can be then used as a \emph{quantum invariant} of $\cM_3$. This feature has been dubbed \emph{3D-3D correspondence} \cite{Dimofte:2011ju,Dimofte:2011py,Cecotti:2011iy}, and continues to be actively investigated \cite{Chung:2014qpa,Dimofte:2014zga,Gukov:2015sna,Pei:2015jsa,Gang:2019uay,Gang:2018hjd,Gang:2017lsr,Alday:2017yxk,Gang:2015wya,Gukov:2017kmk,Eckhard:2018raj,Cheng:2018vpl,Cheng:2022rqr,Chung:2019khu,Cheng:2019uzc,Chun:2019mal,Eckhard:2019jgg,Fan:2020bov}. It is also possible to obtain 3D SCFTs via compactification of 5D SCFTs, see e.g.~\cite{Sacchi:2021afk}.

\section*{Acknowledgments}

We thank Antonio Amariti, Fabio Apruzzi, Cyril Closset, Michele Del Zotto, Yale Fan, Simone Giacomelli, Neil Lambert, Craig Lawrie, Bengt E.W. Nilsson, Carlos Nunez, Shlomo Razamat, Tom Rudelius, Sakura Schafer-Nameki, Jaewon Song, Yuji Tachikawa, Alessandro Tomasiello, Peter West and Gabi Zafrir for helpful comments on the draft, and David Poland and Leonardo Rastelli for spearheading an effort to include hep-th areas in this year's Snowmass process, and inviting us to submit this contribution. 
PCA is supported in part by DOE award DE-SC0011784.
JJH is supported by DOE award DE-SC0013528. 
KI is supported by DOE award DE-SC0009919 and Simons Foundation awards 568420 and 888994. 
MM is supported in part by NSF grants PHY-1915093, by the Simons Foundation grant 815892 and STFC grant ST/T000759/1.

\bibliographystyle{JHEP}

\end{document}